\documentclass[11pt]{article}

\usepackage{amsmath}
\usepackage{graphicx}
\usepackage{indentfirst}
\usepackage{amssymb}
\usepackage{cite}
\usepackage{color}
\usepackage{subfigure}

\begin{document}

\title{\bf{Quasinormal Modes\\in Near-Extremal Spinning C-Metric}}

\date{}
\maketitle

\begin{center}
\author{Bogeun Gwak}$^a$\footnote{rasenis@dgu.ac.kr}\\

\vskip 0.25in
$^{a}$\it{Division of Physics and Semiconductor Science, Dongguk University, Seoul 04620,\\Republic of Korea}\\
\end{center}
\vskip 0.6in

{\abstract
{We investigate the quasinormal modes of the spinning C-metric with a massless scalar field conformally coupled with gravity. Conformal transformation is employed to separate and subsequently solve the massless scalar field equations. As the outer and acceleration horizons approach each other, the potential term of the field equation is reduced to the P\"{o}shl-Teller potential because the acceleration horizon is similar to the cosmological horizon of de Sitter spacetime. Finally, we obtain the analytical quasinormal frequency at which the decay rate is quantized.}}

\thispagestyle{empty}
\newpage
\setcounter{page}{1}

\section{Introduction}\label{sec:01}

Black holes are compact objects with an interesting structure called an event horizon, where no emissions are observed. When matter of any kind passes through the horizon, it is pulled into the singularity at the centre of the black hole. Various solutions have been proposed to characterize black holes. One of the simplest solutions to a static black hole with mass is the Schwarzschild black hole \cite{Schwarzschild:1916uq}. A more realistic solution to a black hole in our universe is the Kerr black hole with angular momentum\cite{Kerr:1963ud}. Ever since the first gravitational wave was detected\cite{LIGOScientific:2016lio}, rotating black holes, such as a Kerr black hole, have been common sources of gravitational waves. However, the most general solution to a black hole with a gauge field is the Pleba\'{n}ski-Demia\'{n}ski black hole. This solution has various parameters that can be set to obtain a well-known form of the black hole\cite{Plebanski:1976gy, Griffiths:2005se, Griffiths:2005qp}. One of the black holes produced by the Pleba\'{n}ski-Demia\'{n}ski solution is the spinning C-metric\cite{Farhoosh:1980zz,Bicak:1999sa,Letelier:1998rx}. This type of black hole comprises two rotating black holes that accelerate in opposite directions. In addition to the spinning C-metric, its exterior and interior can be presumed to those of the accelerating Kerr black hole. Furthermore, the acceleration creates another horizon far from the black hole in spacetime called the acceleration horizon, which plays a similar role to the cosmological horizon in de Sitter spacetime. However, the interesting point is that it appears even if the cosmological constant is zero.

An external observer can investigate various properties of a black hole through its interactions with matter such as particles and waves. In the case of a particle, each black hole possesses its own geodesic to represent its properties. In the case of a wave, the characteristics of a black hole can be determined by studying the scattering of the wave. During wave-scattering, the black hole and wave backreact mutually. In most cases, the energy scale of a black hole is presumed to be larger than that of a wave; therefore, the scattering can be assumed to be a perturbation to a black hole by a wave. As a result of perturbation, the wave can be relaxed, and its oscillations are quasinormal modes (QNMs). Studies on QNMs began with the Schwarzschild black hole\cite{Vishveshwara:1970zz,Press:1971wr} and Kerr black holes \cite{Detweiler:1977gy}. In particular, QNMs can represent the characteristics of black holes based on their frequencies\cite{Kokkotas:1999bd} and can be used to test the stability of a black hole\cite{Brito:2015oca}. Hence, studies on QNMs are central to the research on black holes. Furthermore, given certain information, the frequencies of the QNMs can be obtained in analytical form with a few approximations. One of these is the use of the P\"{o}shl-Teller potential\cite{Poschl:1933zz,Ferrari:1984zz,Beyer:1998nu}. Reducing the potential term of the field equation into the P\"{o}shl-Teller potential, which appears with a positive cosmological constant\cite{Moss:2001ga}, enables us to obtain the exact form of the QNM frequency. Subsequently, the frequencies of the QNMs can be found in Schwarzschild-de Sitter black holes\cite{Cardoso:2003sw} and its higher-dimensional generalizations\cite{Molina:2003ff,Gwak:2019ttv}. Interestingly, QNMs have been studied from various perspectives\cite{Konoplya:2003dd,Konoplya:2007jv,Bini:2008mzd,Dyatlov:2010hq,Dyatlov:2011jd,Kraniotis:2016maw,Khanna:2016yow,Price:2017cjr,Gwak:2018akg,Novaes:2018fry,BarraganAmado:2018zpa,Gwak:2019rcz,Churilova:2021nnc,Gwak:2022mze}. Furthermore, recently studies have highlighted the importance of QNMs in exploring the strong cosmic censorship conjecture\cite{Cardoso:2017soq,Hod:2018lmi,Cardoso:2018nvb,Rahman:2018oso,Dias:2018ynt,Destounis:2019omd,Gim:2019rkl,Miguel:2020uln,Dias:2022oqm,Konoplya:2022kld} and its duality to circular null geodesics\cite{Cardoso:2008bp,Breton:2016mqh,Konoplya:2017wot,Momennia:2019cfd,Guo:2020zmf,Jusufi:2020dhz,Chen:2022ynz,Konoplya:2022gjp}.

In this study, the quasinormal frequency of the spinning C-metric with a massless scalar field conformally coupled with gravity was investigated. In addition, the application of the P\"{o}shl-Teller-potential was extended to spinning black holes with rotation. In fact, the P\"{o}shl-Teller potential cannot appear in black holes with zero cosmological constant due to the absence of the cosmological horizon in de Sitter spacetime. However, an acceleration horizon, which plays a similar role to the cosmological horizon, is present in accelerating black holes. Furthermore, the study was limited to the condition where the outer and acceleration horizons are extremely close together\cite{Dias:2003up}. Assuming this limit to be the near Nariai-type extremal condition in de Sitter black holes, the P\"{o}shl-Teller potential and the analytical form of the quasinormal frequency can be obtained. In other words, the acceleration horizon physically behaves like a cosmological horizon for the QNMs of the scalar field. Furthermore, accelerating black holes have a conformal factor that makes it difficult to separate the field equation because of its complicated form. Therefore, conformal transformation was employed. Considering conformal invariance in the field equation, the scalar field coupled with gravity can be separated by removing the conformal factor\cite{Hawking:1997ia,Destounis:2020pjk}. Finally, the massless scalar field equation can be analytically investigated under near Nariai-type extremal conditions.

The remainder of this paper is organized as follows. In Section \ref{sec:02}, the spinning C metric is reviewed. In Section \ref{sec:03}, the massless scalar field equation is solved by considering conformal transformation. In Section \ref{sec:04}, we the quasinormal frequency under near Nariai-type extremal conditions based on the acceleration horizon is determined. Section \ref{sec:05} summarizes the results.

\section{Spinning C-Metric}\label{sec:02}

The spinning C-metric is considered with electric and magnetic charges under a zero cosmological constant. This is a generalization with the acceleration of the Kerr-Newman (KN) black holes. The spinning C-metric demonstrates a pair of KN black holes moving in opposite directions. Starting from the Pleba\'{n}ski-Demia\'{n}ski metric\cite{Plebanski:1976gy}, the spinning C-metric is given by setting the NUT parameter and cosmological constant to zero. The metric form is\cite{Griffiths:2005se,Griffiths:2005qp}
\begin{align}\label{eq:spinningCmetric01}
ds^2 =\frac{1}{\Omega^2}\left[\frac{Q}{\rho^2}\left(dt-a\sin^2\theta d\phi\right)^2-\frac{\rho^2}{Q}dr^2-\frac{\tilde{P}}{\rho^2}\left(adt-(r^2+a^2)d\phi\right)^2-\frac{\rho^2}{\tilde{P}}\sin^2\theta d\theta^2\right],
\end{align}
and
\begin{align}
&\Omega=1-\alpha r \cos\theta,\\
&\rho^2=r^2+a^2\cos^2\theta,\nonumber\\
&\tilde{P}=\sin^2\theta\left(1-2\alpha M \cos\theta+\alpha^2(a^2+e^2+g^2)\cos^2\theta\right),\nonumber\\
&Q=\left(\left(a^2+e^2+g^2\right)-2Mr+r^2\right)\left(1-\alpha^2 r^2\right),\nonumber
\end{align}
where the metric form corresponds to that considered in \cite{Hong:2004dm}. $m$ is the mass of the black hole, $\alpha$ is the acceleration, $a$ is the spin parameter, and $e$ and $g$ are the electric and magnetic charges, respectively.

The spinning C-metric has three horizons. The inner (Cauchy) and outer (event) horizons are determined using the metric component $Q$ in Eq.\,(\ref{eq:spinningCmetric01}). The locations of the inner and outer horizons are given as
\begin{align}
r_\text{i}=m-\sqrt{m^2-a^2-e^2-g^2},\quad r_\text{h}=m+\sqrt{m^2-a^2-e^2-g^2}.
\end{align}
Another horizon, called the acceleration horizon, is formed from the effect of acceleration. Its location is
\begin{align}
r_\alpha = \alpha^{-1}.
\end{align} 
Furthermore, the locations of the horizons are $r_\text{i}\leq r_\text{h} \leq r_\alpha$, and the observable region of spacetime is $r_\text{h} <r <r_\alpha$. These three horizons provide two types of extremal black holes. First, when the inner and outer horizons coincide, a Kerr-type extremal black hole is formed. Second, when the outer and acceleration horizons coincide, a Nariai-type extremal black hole is formed. In addition, the acceleration horizon plays the role of cosmological horizon even if the cosmological constant is zero.

\section{Solution to Massless Scalar field}\label{sec:03}

Solving the massless scalar field equation is complicated because of the conformal factor of the metric in Eq.\,(\ref{eq:spinningCmetric01}), which hinders the separation of the field equation. However, the massless scalar field equation can be considered as a conformally coupled equation with gravity, given by
\begin{align}\label{eq:scalareq01}
\frac{1}{\Psi}\Box_g \Psi-\frac{1}{6}R=0,\quad R=0,
\end{align}
which is conformally invariant but physically identical to the massless scalar field equation\cite{Destounis:2020pjk}. Subsequently, we perform the conformal transformations as follows
\begin{align}
\tilde{g}_{\mu\nu}\rightarrow\Omega^2 g_{\mu\nu},\quad \tilde{\Psi}\rightarrow \Omega^{-1}\Psi,
\end{align}
which eliminates the conformal factor from the spinning C-metric.
\begin{align}\label{eq:spinningCmetric02}
d\tilde{s}^2&=-\frac{Q}{\rho^2}\left(dt-a\sin^2\theta d\phi\right)^2+\frac{\rho^2}{Q}dr^2+\frac{\rho^2}{\Delta_\theta}d\theta^2+\frac{\Delta_\theta \sin^2\theta}{\rho^2}\left(adt-(r^2+a^2)d\phi\right)^2,
\end{align}
with
\begin{align}
\tilde{P}&=\sin^2\theta \Delta_\theta(\theta),
\end{align}
where no conformal factors exist in the metric component. Note that the locations of the horizons are identical to those of the previous horizons. Using the conformally scaled metric in Eq.\,(\ref{eq:spinningCmetric02}), the field equation in Eq.\,(\ref{eq:scalareq01}) becomes
\begin{align}\label{eq:scalareq02}
\frac{1}{\tilde{\Psi}}\Box_{\tilde{g}} \tilde{\Psi}-\frac{1}{6}\tilde{R}=0,
\end{align}
and 
\begin{align}
\tilde{R}=-\frac{1}{\rho^2}(-2\Delta_\theta+3\cot\theta\Delta_\theta'+Q''+\Delta_\theta'').
\end{align}
The solution to the scalar field equation in Eq.\,(\ref{eq:scalareq01}) can be obtained from that of the conformally scaled version in Eq.\,(\ref{eq:scalareq02}). Thus, a solution of Eq.\,(\ref{eq:scalareq02}) represents the massless scalar field in the spinning C-metric. Furthermore, due to the absence of a conformal factor in the metric in Eq.\,(\ref{eq:scalareq02}) the field equation becomes considerably simpler.  

The scalar field equation with conformal coupling is expressed as
\begin{align}
\frac{1}{\sqrt{-\tilde{g}}}\partial_\mu \left(\sqrt{-\tilde{g}}\tilde{g}^{\mu\nu}\partial_\nu \tilde{\Psi}\right)-\frac{1}{6}\tilde{R}\tilde{\Psi}=0,\quad \text{det}\tilde{g}=-\rho^4\sin^2\theta,\quad \sqrt{-g}=\rho^2\sin\theta.
\end{align}
Considering the translational symmetries of the conformally scaled metric to the time and $\theta$ directions, the ansatz for the conformally scaled scalar field can be given as
\begin{align}
\tilde{\Psi}(t,r,\theta,\phi)=e^{-i\omega t} e^{im\phi} \psi(r) \Theta(\theta).
\end{align}
The frequency and quantum number of $\phi$-directional angular momentum are denoted by $\omega$ and $m$, respectively. Under the ansatz, the scalar field equation is clearly separable. By introducing the separate variable $\mathcal{K}$, we obtain the radial and $\theta$-directional equations. Then, the radial equation is
\begin{align}\label{eq:thetaequation02}
\partial_r(Q\partial_r \psi(r))+\left(\frac{\left(\omega(r^2+a^2)-am\right)^2}{Q}+\frac{1}{6}Q''-\mathcal{K}\right)\psi(r)=0,
\end{align}
and the $\theta$-directional equation is
\begin{align}\label{eq:thetaequation03}
\frac{1}{\sin\theta \Theta(\theta)}\partial_\theta(\sin\theta \Delta_\theta\partial_\theta\Theta(\theta))-\frac{a^2 \omega^2\sin^2\theta}{\Delta_\theta}+\frac{2a m\omega}{\Delta_\theta}
-\frac{m^2\csc^2\theta}{\Delta_\theta}+\frac{1}{6}(-2\Delta_\theta+3\cot\theta\Delta_\theta'+\Delta_\theta'')+\mathcal{K}=0.
\end{align}
The details of the modes are given in the frequency $\omega$ of the radial equation. Furthermore, the continuous condition of the $\theta$-directional solution ensures that the ansatz well works. Using the transformation, we get
\begin{align}
\frac{dz}{d\theta}=\frac{1}{\sin\theta \Delta_\theta},
\end{align}
Subsequently, the $\theta$-directional equation in Eq.\,(\ref{eq:thetaequation03}) can be expressed in terms of $z$
\begin{align}\label{eq:thetaequation04}
\frac{1}{\Theta(z)}\partial^2_z \Theta(z)-m^2+\sin^2\theta\left(-a^2 \omega^2\sin^2\theta+2a m\omega+\frac{1}{6}\Delta_\theta(-2\Delta_\theta+3\cot\theta\Delta_\theta'+\Delta_\theta'')+\mathcal{K}\Delta_\theta\right)=0.
\end{align}
The $\theta$-directional boundaries are $0$ and $\pi$ which correspond to $\mp\infty$ in the $z$-direction. Then, Eq.\,(\ref{eq:thetaequation04}) can be rewritten as
\begin{align}\label{eq:boundary1001}
\frac{1}{\Theta(z)}\partial^2_z \Theta(z)-m^2=0.
\end{align}
Based on the above Eq.\,(\ref{eq:boundary1001}), the solutions at the boundaries are
\begin{align}
\Theta(z)&\sim e^{+mz},\,\,z\rightarrow -\infty\,\,(\theta\rightarrow 0),\\
\Theta(z)&\sim e^{-mz},\,\,z\rightarrow +\infty\,\,(\theta\rightarrow \pi),\nonumber
\end{align}
However, these solutions are finite at the boundary, and these forms in boundaries are the similar to the C-metric without a spinning\cite{Destounis:2020pjk}. Furthermore, the $\theta$-directional solutions are continuous at the boundary, which ensures that the ansatz is physically meaningful.

The radial solution is associated with the energy of the scalar field or decay rate; therefore, it is essential in understanding the characteristics of the black hole. The radial equation becomes the Schr\"{o}dinger-type equation under the tortoise coordinate given by
\begin{align}
\frac{dr^*}{dr}=\frac{r^2+a^2}{Q},\quad \psi(r)\rightarrow\frac{\tilde{\psi}(r)}{\sqrt{r^2+a^2}}.
\end{align}
The rewritten radial equation is obtained as
\begin{align}\label{eq:radialstypeeq03}
\frac{d^2 \tilde{\psi}(r^*)}{d{r^*}^2}+\left(\left(\omega-m\Omega\right)^2+\frac{QQ''}{6(r^2+a^2)^2}-\frac{\mathcal{QK}}{(r^2+a^2)^2}-\frac{r Q'Q}{(r^2+a^2)^3}-\frac{Q^2(a^2-2r^2)}{(r^2+a^2)^4}\right)\tilde{\psi}(r^*)=0,
\end{align}
which includes information on the radial behaviors of the scalar field when the equation is numerically solved in all ranges. However, our main focus is on near-horizon behavior; therefore, the radial equation in Eq.\,(\ref{eq:radialstypeeq03}) is approximated to obtain the near-horizon regions for the $Q(r)\rightarrow 0$, $r\rightarrow$ $r_\text{h}$ and $r_\alpha$. Under this limit, the radial equation becomes
\begin{align}
\frac{d^2\tilde{\psi}}{{dr^*}^2}+\left(\omega-m\Omega_\text{h}\right)^2\tilde{\psi}=0, \quad \frac{d^2 \tilde{\psi}}{{dr^*}^2}+\left(\omega-m\Omega_\alpha\right)^2\tilde{\psi}=0,
\end{align}
where
\begin{align}
\lim_{r\rightarrow r_\text{h}}\Omega=\Omega_\text{h},\quad \lim_{r\rightarrow r_\alpha}\Omega=\Omega_\alpha. 
\end{align}
Then, the solutions are
\begin{align}
\tilde{\psi}(r\rightarrow r_\text{h})&=\mathcal{T}e^{-i(\omega-m\Omega_\text{h})r^*},\quad\text{(ingoing)}\\
\tilde{\psi}(r\rightarrow r_\alpha)&=\mathcal{O}e^{+i(\omega-m\Omega_\alpha)r^*},\quad\text{(outgoing)}\nonumber
\end{align}
where the wave enters both horizons only at the boundary condition. However, the frequencies representing the oscillation and decay rates of scalar modes also need to be determined.

\section{QNMs in Near Nariai-Type Extremal Spinning C-Metric}\label{sec:04}

QNMs are crucial to the investigation of black holes. In particular, analytical analysis reveals relationships of the QNMs with the parameters of a black hole. In this study, near Nariai-type extremal black holes are considered, in which {\it{the outer and acceleration horizons are close to each other}}. Moreover, the scalar field with black holes produces a P\"{o}shl-Teller potential, for which the QNMs have analytic solutions.

\begin{figure}[h]
\centering
\subfigure[{Phases in $(\beta,\alpha)$ with $M=1$.}] {\includegraphics[scale=0.9,keepaspectratio]{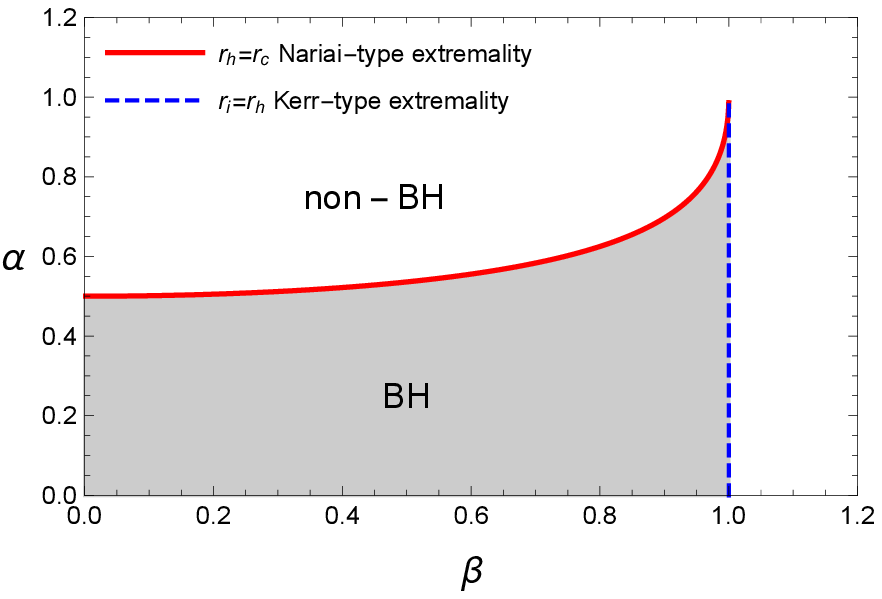}}\quad
\subfigure[{Phases in $(M,\alpha)$ with $\beta=1$.}] {\includegraphics[scale=0.9,keepaspectratio]{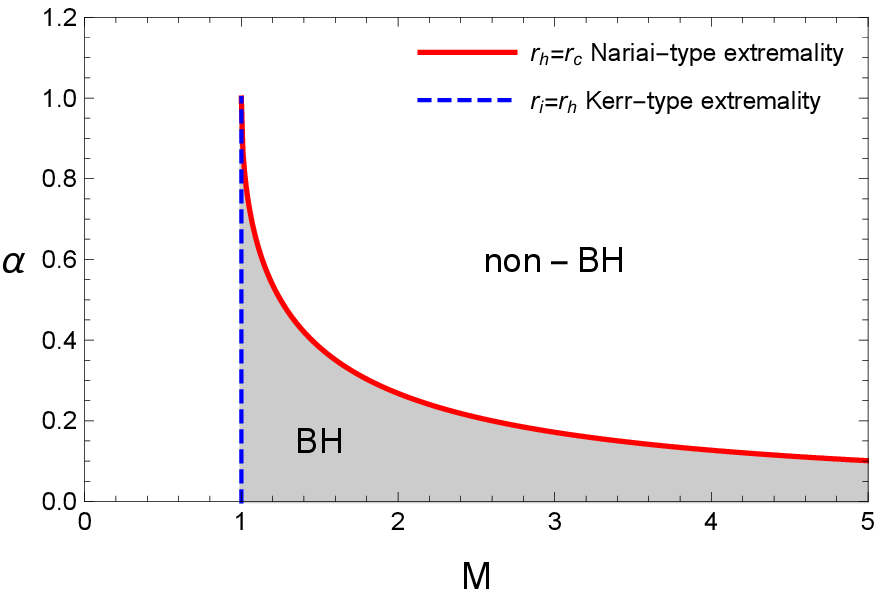}}
\caption{{\small The extremal conditions of spinning C-metric by $(M,\alpha,\beta)$.}}
\label{fig:phases01}
\end{figure}

The parameter regions can be limited by two extremal conditions. The parameters $a$, $e$, and $g$ are equally contributed to the extremal condition, so $\beta^2\equiv a^2+e^2+g^2$, shortly. The Kerr-type extremal condition is $r_\text{i}=r_\text{h}$, and the Nariai-type extremal condition is $r_\text{h}=r_\alpha$
\begin{align}
M=\beta,\quad \alpha=\frac{1}{M+\sqrt{M^2-\beta^2}}.
\end{align}
Furthermore, there is a condition to make $r_\text{i}=r_\text{h}=r_\alpha$: $\alpha=M^{-1}$. This is shown as Fig.\ref{fig:phases01}. 
Then, our investigation is focued on the near Nariai-type extremal condition which is near to the red line in Fig.\ref{fig:phases01}.

The near Nariai-type extremal condition can be summarized as $r_\text{h}\approx r_\alpha$\cite{Gwak:2019ttv}. Under the condition that the outer and acceleration horizons are extremely close, the radial coordinate is limited to $r_\text{h}\leq r \leq r_\alpha$ where the metric component is
\begin{align}
Q\ll 1, \quad Q'\ll 1.
\end{align}
Considering the first orders of functions $Q$ and $Q'$, the Schr\"{o}dinger-type radial equation in Eq.\,(\ref{eq:radialstypeeq03}) under the near Nariai-type extremal condition is obtained as
\begin{align}\label{eq:radialnearntextremal15}
\frac{d^2 \tilde{\psi}}{{dr^*}^2}+\left(\left(\omega-m\Omega_\text{}\right)^2+\frac{Q Q''}{6(r^2+a^2)^2}-\frac{\mathcal{K}Q}{(r^2+a^2)^2}\right)\tilde{\psi}=0.
\end{align}
Note that the range of the radial coordinates is limited under the near Nariai-type extremal condition. Thus, assuming that
\begin{align}
r\approx r_\text{h}\approx r_\alpha.
\end{align}
The metric component and its derivative are closely associated. Function $Q$ can have four solutions including $r_\text{h}$ and $r_\alpha$. These can be rewritten as:
\begin{align}
Q=\alpha^2(r-r_\text{h})(r_\alpha-r)(r-r_1)(r-r_2),\quad  \left.Q'\right|_{r=r_\text{h}}\equiv Q'_\text{h}=\alpha^2 (r_\alpha-r_\text{h})(r_\text{h}-r_1)(r_\text{h}-r_2),
\end{align}
where $r_1$ and $r_2$ are also solutions; however, they can be complex or lie outside the physical range of the radial coordinate. Because function $Q'$ is proportional to the surface gravity, the function $Q$ can be rewritten in terms of surface gravity as \cite{Debnath:2015tda,Anabalon:2018qfv}
\begin{align}
Q=\frac{2(r_\text{h}^2+a^2)\kappa_\text{h}}{r_\alpha-r_\text{h}}(r-r_\text{h})(r_\alpha-r), \quad \kappa_\text{h}=\frac{Q'_\text{h}}{2(r_+^2+a^2)}.
\end{align}
Using the tortoise coordinate, the radial coordinate can also be approximated under extremal condition as
\begin{align}
r&=\frac{r_\text{h}+r_\alpha e^{2\kappa_\text{h} r^*}}{1+e^{2\kappa_\text{h} r^*}}.
\end{align}
Then, the function $Q$ becomes
\begin{align}
Q=\frac{(r_\alpha-r_\text{h})(r_\text{h}^2+a^2)\kappa_\text{h}}{2\cosh^2{\kappa_\text{h}r^*}}.
\end{align}
Because function $Q$ appears in the effective potential in Eq.\,(\ref{eq:radialnearntextremal15}), the radial equation can be rewritten at the Nariai-type extremal limit.
\begin{align}\label{eq:radialeq115}
&\frac{d^2R}{{dr^*}^2}+\left(\left(\omega-m\Omega_\text{h}\right)^2-\frac{V_0}{\cosh^2{\kappa_\text{h}r^*}}\right)R(r)=0,
\end{align}
and
\begin{align}
V_0=\frac{1}{2}\kappa_\text{h}\left(\frac{r_\alpha-r_\text{h}}{r_\text{h}^2+a^2}\right)\left(\frac{Q''_\text{h}}{6}-\mathcal{K}\right),\quad Q''_\text{h}=2-2\alpha^2(a^2+e^2+g^2+6r_\text{h}(r_\text{h}-m)).
\end{align}
The radial equation in Eq.\,(\ref{eq:radialeq115}) has the P\"{o}shl-Teller potential where the general solution of the frequency is a well- known complex form. Subsequently, the real part of the frequency is obtained as
\begin{align}
\text{Re}(\omega)=m\Omega_\text{h}+\kappa_\text{h}\sqrt{\frac{1}{2\kappa_\text{h}}\left(\frac{r_\alpha-r_\text{h}}{r_\text{h}^2+a^2}\right)\left(\frac{Q''_\text{h}}{6}-\mathcal{K}\right)-\frac{1}{4}}.
\end{align}
The imaginary part is associated with the scalar field decay rate.
\begin{align}
\text{Im}(\omega)=-\left(n+\frac{1}{2}\right)\kappa_\text{h},\quad n=0,1,2...
\end{align}
In the imaginary part of the frequency, a negative value implies that the scalar field decays with respect to the evolution of time. The decay rate is quantized using an integer $n$. Furthermore, the higher $n$ modes decay rapidly owing to the large decay rate; therefore, the most dominant mode is $n=0$, which has the lowest decay rate and is proportional to the surface gravity.

\section{Summary}\label{sec:05}

The QNMs of the massless scalar field in the spinning C-metric were investigated under the Nariai-type near-extremal condition. However, a significant obstacle was present; the spinning C-metric has a conformal factor depending on the coordinates; therefore, its massless scalar field equation is not separable. Accordingly, the field equation was considered at the point where the scalar field is conformally coupled with gravity and separated using conformal invariance.

Furthermore, despite the absence of the cosmological constant, the spinning C-metric could exist in the Nariai-type near-extremal condition, where the outer and acceleration horizons are close to each other, instead of the cosmological horizon. This near-extremality enabled us to obtain the QNMs analytically under these conditions. Subsequently, the separated radial equation were rewritten in terms of tortoise coordinates under near-extremal conditions. In particular, the potential term became the P\"{o}shl-Teller potential with an analytical solution with quantized decay rates. In addition, the real part of the frequency was found to depend on the parameters of the spinning C-metric, including the acceleration factor, whereas the imaginary part, representing the decay rate, depended only on the surface gravity. Furthermore, the dominant QNM of the scalar field was found to be the $n=0$ mode, which has the lowest decay. Therefore, due to the acceleration horizon, the spinning C-metric behaves like a black hole with a positive cosmological constant, and the QNM with $n=0$ is the most dominant mode.

\vspace{10pt} 

\noindent{\bf Acknowledgments}

\noindent This work was supported by the National Research Foundation of Korea (NRF) grant funded by the Ministry of Science and ICT (NRF-2018R1C1B6004349) and the Ministry of Education (NRF-2022R1I1A2063176) and the Dongguk University Research Fund of 2022. BG appreciates APCTP for its hospitality during the topical research program, {\it Multi-Messenger Astrophysics and Gravitation}.

\bibliography{References_v2}
\bibliographystyle{Refsty}
\end{document}